\documentclass[lettersize,journal]{IEEEtran}
\IEEEoverridecommandlockouts
\usepackage[utf8]{inputenc}
\usepackage{cite,balance}
\usepackage{amsmath,amssymb,amsfonts,multirow,upgreek,subcaption}
\usepackage[table]{xcolor}
\usepackage{textcomp}
\usepackage{xcolor}
\usepackage{bm}
\usepackage{algpseudocode}
\usepackage{algorithm}
\usepackage{enumitem}
\usepackage{psfrag}
\usepackage{tikz}
\usetikzlibrary{shapes.geometric, arrows}
\tikzstyle{roundrec} = [font=\footnotesize,rectangle, rounded corners, minimum width=1cm, minimum height=1cm,text centered, text width=2.5cm, draw=black]
\tikzstyle{dotted_roundrec} = [font=\footnotesize,rectangle, rounded corners, minimum width=1cm, minimum height=1cm,text centered, text width=2.5cm, draw=black,dashed]
\tikzstyle{rec} = [font=\footnotesize,rectangle, minimum width=3cm, minimum height=1cm, text width=6cm, draw=black]
\tikzstyle{long_rec} = [font=\footnotesize,rectangle, minimum width=3cm, minimum height=1cm, text width=8.75cm, draw=black]
\tikzstyle{bubble} = [font=\footnotesize,rectangle, rounded corners, text centered, draw=black]
\tikzstyle{arrow} = [thick,->,>=stealth]
\usepackage{titlesec}

\def\BibTeX{{\rm B\kern-.05em{\sc i\kern-.025em b}\kern-.08em
    T\kern-.1667em\lower.7ex\hbox{E}\kern-.125emX}}

\renewcommand{\i}{\mathrm{i}}
\renewcommand{\j}{\mathrm{j}}
\DeclareMathOperator*{\argmax}{arg\,max}

\newcommand*{\herm}{^{\mathsf{H}}}
\newcommand*{\transp}{^{\mathsf{T}}}

\titlespacing\section{-2pt}{12pt plus 4pt minus 2pt}{0pt plus 2pt minus 2pt}
\titlespacing\subsection{-1pt}{12pt plus 4pt minus 2pt}{0pt plus 2pt minus 2pt}

\title{Joint User Detection and Localization in Near-Field Using Reconfigurable Intelligent Surfaces}

\author{Georgios Mylonopoulos, Behrooz Makki, \emph{Senior Member IEEE}, Stefano Buzzi, \emph{Senior Member IEEE}, G\'abor Fodor, \emph{Senior Member IEEE} 
\thanks{This work has received funding from the European Union’s Horizon 2020 research and innovation programme under the Marie Skłodowska-Curie grant agreement No 956256. S. Buzzi was also supported by the European Union under the Italian National Recovery and Resilience Plan (NRRP) of NextGenerationEU, partnership on “Telecommunications of the Future” (PE00000001 - program “RESTART”, Structural Project SRE). \\
G. Mylonopoulos and S. Buzzi are with the Department of Electrical and Information Engineering, University of Cassino and Southern Lazio, 03043 Cassino, Italy, and also with the Consorzio Nazionale Interuniversitario per le Telecomunicazioni (CNIT), 43124 Parma, Italy (e-mail: georgios.mylonopoulos / buzzi@unicas.it). S. Buzzi is also with Politecnico di Milano, Milan, Italy. \\
B. Makki and G.Fodor are with Ericsson Research, Sweden (e-mail: behrooz.makki / gabor.fodor@ericsson.com). G. Fodor is also with the Division of Decision and Control, KTH Royal Institute of Technology, 11428 Stockholm, Sweden.}
}
\date{\today}

\begin{document}

\maketitle

\begin{abstract}
This letter studies the problem of jointly detecting active user equipments (UEs) and extracting position information in the near field, wherein the base station (BS) is unaware of the number of active (or inactive) UEs and their positions, nor is able to pre-assign pilot signals. The system is equipped with multiple passive reconfigurable intelligent surfaces (RISs) that provide a low-complexity solution for detection and localization, with additional degrees of freedom due to the additional inspection points. Specifically, we propose an iterative detection procedure, allowing the BS to assign pilots to detected UEs, thus providing a structured channel access. Also, the problems of multiple access interference and multipath suppression are explored as limiting performance factors. The results show that, with a proper implementation of the RISs, our proposed scheme can detect and localize the UEs, augmenting benchmark UE detection schemes to spatially aware detection.
\end{abstract}


\section{Introduction}
As the number of connected user equipment (UE) devices increases, wireless networks are expected to support massive connectivity for Internet-of-Things and machine-type communication applications. In such scenarios, a large number of UEs may be connected to a base station (BS), but the device activity is typically sporadic and only a limited number of UEs may be active at a given point\cite{9205230}. Therefore, the active UEs may be initially unknown to the BS for a structured channel access, and effective UE detection techniques are needed to establish communication links. In that sense, activity detection and the effect of geometry and device density are explored in~\cite{zhang2023grant}, while~\cite{li2022joint}, jointly investigates channel estimation and UE activity detection in massive access networks.

Spatial awareness is a vital sensing feature for many practical applications and novel positioning techniques need to be derived for integrated sensing and communication (ISAC)~\cite{Wei:22}. As the number of UEs increases, the probability that the BS cannot detect/track them increases due to, e.g., blockages and \emph{localization holes}, where the BS alone cannot detect them accurately. In these cases, the presence of reconfigurable intelligent surfaces (RISs) can improve the detection/localization accuracy as they provide a \emph{second view} and bypass blockages. 

RISs have gained much attention in ISAC applications, as they provide partial control over the propagation environment and enhance signal reception~\cite{10149664}. RISs of various architectures have been considered in dedicated localization schemes~\cite{10621044,dardari2021nlos} and in radar applications~\cite{9732186}. In RIS aided systems localization may be an auxiliary application to some other procedure, e.g., joint localization and channel estimation~\cite{10149471}. In spatially aware applications, RISs demonstrate more degrees of freedom in the near-field (NF), due to the the spherical nature of the reflected wavefronts~\cite{dardari2021nlos,9732186,10149471}. Moreover, RIS-assisted UE detection has been investigated in the far-field (FF)~\cite{10117546}. However, to the best of our knowledge, the problem of jointly detecting active UEs and extracting positional information in the NF has not been investigated.

In this letter we study the problem of UE activity detection, where the large and unknown number of active UEs prohibits the BS of pre-assigning pilot sequences. The active UEs proceed with unstructured and random channel access, leading to collisions and inter-user interference among them. As the BS tackles the problem of signal detection, we propose that positional information may be extracted, for the active UEs, without a dedicated localization procedure. Moreover, we investigate the versatility of RISs in augmenting the sensing capabilities of existing infrastructure; exploiting the NF characteristics of the considered scenario, we design a point-wise, RIS-assisted, scanning procedure. In that sense, we provide an extended discussion on the practical advantages of RIS-aided inspection and present an intuitive guide for practical deployment and configuration of RISs for sensing applications. We propose an iterative procedure, where the random multipath interference is suppressed and the BS gradually allocates resources among the detected UEs, reducing the channel congestion. Also, we compare the performance of the proposed scheme with a benchmark UE detection scheme with no RIS and no spatial awareness. Our results show that, our proposed iterative scheme can localize the NF UEs with high accuracy.

\section{System Model \& Detection Procedure}
\begin{figure}[t]
    \centering
    \includegraphics[width=\columnwidth]{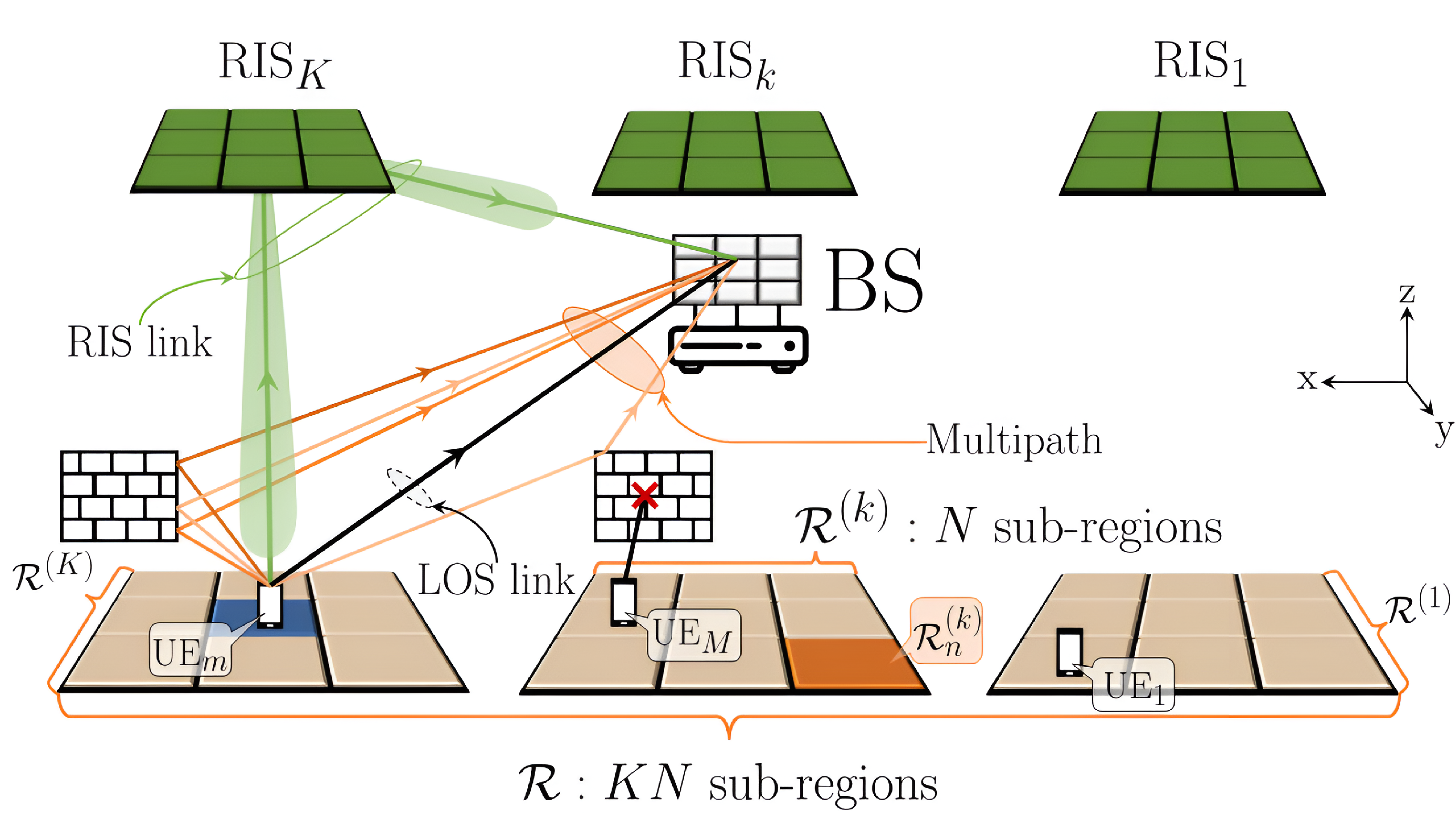}
    \caption{The BS detects and localizes the activity of $M$ UEs, with the aid of $K$ RISs; $N$ different RIS configurations perform a scanning procedure as the UEs make $N$ transmissions.}
    \label{fig:sm}
\end{figure}

We consider the joint problem of detecting UE activity and estimating their position. A set of $K$ passive RISs undertake a scanning procedure, correlating the detected signals with the position of their source, i.e., the UEs. Fig.~\ref{fig:sm} illustrates the considered setup. There is a BS, wall-mounted on the x-z plane, at point $\bm{p}_{\rm{BS}}$, equipped with a uniform planar array (UPA) of $N_{\rm BS}$ elements, i.e., $N_{\rm{BS}} = N_{\rm{BS}}^{(\rm{x})}\times N_{\rm{BS}}^{(\rm{z})}$ with an antenna spacing of $\delta_{\rm{BS}}$, across both dimensions. There are $M$ single antenna UEs, with unknown positions; $\bm{p}_{\rm{UE}}^{(m)} \in \mathcal{R} ,m=1,\dots,M$, where $\mathcal{R}$ is a 3-dimensional region. The number of UEs, $M$, is not known to the BS. While we present the setup for the general case with multiple RISs, depending on the considered area and quality-of-service requirements, a single RIS may be considered. Each RIS is ceiling-mounted on the x-y plane, at a fixed position, $\bm{p}_{\rm{RIS}}^{(k)} ,k=1,\dots,K$, known to the BS, equipped with $N_{\rm{RIS}}^{(k)}$ elements, i.e., $N_{\rm{RIS}}^{(k)} = N_{\rm{RIS}}^{(k,\rm{x})}\times N_{\rm{RIS}}^{(k,\rm{y})}$, with an element spacing of $\delta_{\rm{RIS}}^{(k)}$. Each RIS is designed to aid with the detection of UEs in a smaller region, i.e., the smaller region inspected by the $k^{\rm{th}}$ RIS is denoted by $\mathcal{R}^{(k)} \in \mathcal{R}$.

\subsection{Detection Procedure}
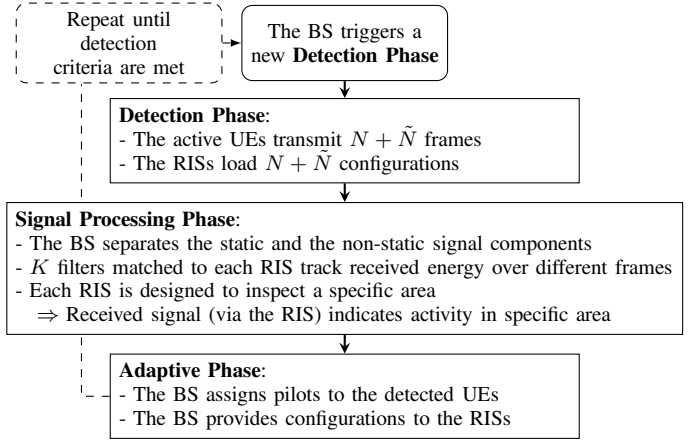
\begin{figure}[t]
    \centering
    \begin{tikzpicture}[node distance=2cm]
        \node (start) at (0,0) [roundrec] {The BS triggers a new \textbf{Detection Phase}};
        \node (det_phase) at (0,-1.3) [rec] {\textbf{Detection Phase}:\\- The active UEs transmit $N+\tilde{N}$ frames\\- The RISs load $N+\tilde{N}$ configurations};
        \node (sign_phase) at (0,-3) [long_rec] {\textbf{Signal Processing Phase}:\\- The BS separates the static and the non-static signal components\\- $K$ filters matched to each RIS track received energy over different frames\\- Each RIS is designed to inspect a specific area\\$\quad\Rightarrow$ Received signal (via the RIS) indicates activity in specific area};
        \node (adapt_phase) at (0,-4.7) [rec] {\textbf{Adaptive Phase}:\\- The BS assigns pilots to the detected UEs\\- The BS provides configurations to the RISs};
        \node (rep) at (-3,0) [dotted_roundrec] {Repeat until detection criteria are met};

        \draw [arrow] (node cs:name=start,anchor=south) -- (node cs:name=det_phase,anchor=north);
        \draw [arrow] (node cs:name=det_phase,anchor=south) -- (node cs:name=sign_phase,anchor=north);
        \draw [arrow] (node cs:name=sign_phase,anchor=south) -- (node cs:name=adapt_phase,anchor=north);
        \draw[dashed,-latex] (node cs:name=rep,anchor=east) -- (node cs:name=start,anchor=west);
        \draw[dashed] (node cs:name=adapt_phase,anchor=west) --(-3.5,-4.7)--(-3.5,-3.9);
        \draw[dashed] (-3.5,-2.05)--(-3.5,-0.55);
    \end{tikzpicture}
    \caption{Proposed UE detection protocol with spatial awareness, multipath suppression and adaptive resource allocation.}
    \label{fig:det_proc}
\end{figure}

Our proposed solution combines three key ideas that have not been considered before within the UE detection framework, as illustrated in Fig.~\ref{fig:det_proc}. The BS broadcasts a signal, synchronizing the UEs that proceed to a set of transmissions. For the commonly considered OFDM structure, the available subcarriers and timeslots create $B$ orthogonal pilots, $\bm{x}_{q}^{(b)} = [x_{q}^{(1,b)},\dots,x_{q}^{(L,b)}]\transp \in \mathbb{R}^{L\times 1}$, of $L$ timeslots for $q = 1,\dots,Q,$ subcarriers, with $B=LQ$ and a subcarrier spacing of $W_{\rm{sub}}$. The pilots are also referred to as \emph{resource blocks} (RBs). Each UE autonomously and randomly picks a pilot sequence from a pre-determined set, leading to possible collisions. If a UE has been detected, an RB is assigned by the BS, for a reliable, collision-free procedure. Note that traditional uplink localization or estimation schemes require a \emph{known} pilot for each UE, e.g.,~\cite{10621044,10149471}.

We propose to split the pilots in two disjoint sets: the set of pilots to be randomly picked by undetected UEs, of cardinality $B_{\rm{R}}$, and the set of pilots assigned by the BS during the \emph{adaptive phase}, of cardinality $B_{\rm{A}}$. The UEs transmit $\tilde{N} + N$ OFDM frames using their respective pilots, where $\tilde{N}$ frames are used during the \emph{signal processing phase} to separate the RIS-assisted link and suppress the multipath component. For the $N$ transmissions, the $k^{\rm{th}}$ RIS uses $N$ configurations designed to scan $\mathcal{R}^{(k)}$. Thus, $\mathcal{R}^{(k)}$ forms $N$ sub-regions, $\mathcal{R}_{n}^{(k)}$ for $n=1,\dots,N$. Each sub-region is individually inspected during each OFDM frame. The process is iterated periodically, as designed by the network designer.

\subsection{Near-Field Channel \& Signal Model}
The channel consists of two main components: \emph{static} and \emph{non-static}. The static component, $\dot{\bm{h}}$, refers to all links that do not vary, during a single detection phase and encapsulates both the line-of-sight (LOS) link with the BS and the multipath generated by an unknown number of scatterers. The static link between the BS and the $m^{\rm{th}}$ UE, for the $q^{\rm{th}}$ subcarrier, is 
\begin{align}\label{eq:channel_los}
\dot{\bm{h}}_{q}^{(m)} &= \frac{\dot{\beta}_{q}^{(m)}}{\sqrt{K_{\rm{Rice}}^{(m)}+1}} \biggl[\psi^{(m)}\sqrt{K_{\rm{Rice}}^{(m)}}\bm{\alpha}_{q}(\bm{p}_{\rm{UE}}^{(m)}) + \bm{m}_{q}^{(m)}\biggr] ,
\end{align}
for $m = 1,\dots,M$, and $q = 1,\dots,Q$, where $K_{\rm{Rice}}$ is the Rice factor, $\psi\in\{0,1\}$ is a binary indicator for a LOS link and $\bm{\alpha}_{q}(\bm{p})$ is the BS's NF steering vector for position $\bm{p}$. Essentially, $\dot{\bm{h}}_{q}$ is a Rayleigh (Rice) channel if the LOS is (not) obstructed. For the $q^{\rm{th}}$ subcarrier, the NF steering vector is
\begin{align}\label{eq:alpha_NF}
\bigl[\bm{\alpha}_{q}(\bm{p})\bigr]_{\i} &= G(\bm{\theta}_{\bm{p},\i})\frac{d_{\rm{b}}}{d_{\i,\rm{b}}}\text{exp}\bigl\{-j2\pi\frac{d_{\i,\rm{b}}}{\lambda_{q}}\bigr\} \;,
\end{align}
where $d_{\i,\rm{b}}$ is the distance from the $\i^{\rm{th}}$ antenna to point $\bm{p}$ and $G(\bm{\theta})$ is the scalar element gain for the angle of arrival (or departure), $\bm{\theta} = [\theta_{\rm{az}}, \theta_{\rm{el}}]^{T}$, for each element-to-element link. In~\eqref{eq:channel_los}, $\bm{m}\sim\mathcal{CN}(\bm{0},\bm{I})$ denotes the random multipath and the scalar $\dot{\beta}$ is the channel's complex amplitude, given by
\begin{align}
\dot{\beta}_{q}^{(m)} &= \frac{\lambda_{q}}{4\pi d_{\rm{b}}^{(m)}}\text{exp}\bigl\{j\phi_{o}^{(m)}\bigr\} \;,
\end{align}
where $\lambda_{q}$ is the wavelength, $d_{\rm{b}} = \|\bm{p}_{\rm{UE}} - \bm{p}_{\rm{BS}}\|$ and $\phi_{o} \in [0,2\pi]$  accounts for the synchronization mismatch. 

Finally, the non-static component, $\bar{\bm{h}}$, refers to all links between the $m^{\rm{th}}$ UE and the BS via the RISs, as they vary with each RIS configuration. For the $q^{\rm{th}}$ subcarrier, we have
\begin{align}
\bar{\bm{h}}_{q,n}^{(m)} = \sum_{k=1}^{K} \bm{H}_{k,q}\text{diag}\{\bm{\omega}_{n}^{(k)}\}\bm{r}_{k,q}^{(m)}
\end{align}
where $\bm{\omega}$ marks the RIS configuration, $\bm{r}$ denotes the channel between each UE and the RIS and $\bm{H}$ denotes the channel between the RIS and the BS. The channels involving the $m^{\rm{th}}$ UE, the $\i^{\rm{th}}$ element of the $k^{\rm{th}}$ RIS and $\j^{\rm{th}}$ BS antenna are
\begin{subequations}
\begin{align}
\bigl[ \bm{r}_{k,q}^{(m)}\bigr]_{\i} =&  \frac{G(\bm{\theta}_{k,\i}^{(m)})\lambda_{q}}{4\pi d_{\i,\rm{r}}^{(k,m)}}\text{exp}\bigl\{-j2\pi\frac{d_{\i,\rm{r}}^{(k,m)}}{\lambda_{q}} + j\phi_{o}^{(m)}\bigr\} \;, \label{seq:channel_ru}\\
\bigl[ \bm{H}_{k,q}\bigr]_{\i,\j} =&  \frac{G(\bm{\theta}_{k,\i})G(\bm{\theta}_{k,\j})\lambda_{q}}{4\pi d_{\i,\j}^{(k)}}\text{exp}\bigl\{-j2\pi\frac{d_{\i,\j}^{(k)}}{\lambda_{q}}\bigr\} \;,\label{seq:channel_br}
\end{align}
\end{subequations}
where $d_{\i,\rm{r}}^{(k,m)}$ and $d_{\i,\j}^{(k)}$ denote the distance from the $\i^{\rm{th}}$ RIS element to the $m^{\rm{th}}$ UE and the $\j^{\rm{th}}$ BS antenna, respectively. Note that $\bar{\bm{h}}_{q,n}$ is non-static, as it varies among different frames. 

The received signal during the $n^{\rm{th}}$ OFDM frame is
\begin{align}
\bm{y}_{q,n} = \sqrt{P_{\rm{sym}}} \sum_{m=1}^{M}\bm{h}_{q,n}^{(m)}\otimes \bm{x}_{q,n}^{(m)} + \bm{z}_{q,n} \in \mathbb{C}^{(N_{\rm{BS}} L)\times 1}\; ,
\end{align}
for $q = 1,\dots,Q$, where $P_{\rm{sym}}$ is the transmit power per symbol, $\bm{h}_{q,n}^{(m)} = \dot{\bm{h}}_{q}^{(m)} + \bar{\bm{h}}_{q,n}^{(m)}$ and $\bm{z}_{q,n}$ is a circularly-symmetric Gaussian-random vector $\sim\mathcal{CN}\left(\bm{0},\sigma_{z}^{2}\bm{I}\right)$.

\begin{algorithm}[h!]
\caption{RIS Configuration Design for inspected points}\label{alg:omega_design}
\begin{algorithmic}[1]
\State Set design parameters for the penalty functions, i.e., \newline 
$\beta_{\rm{ps}}=0.1,\beta_{\rm{SL}}=0.1,\alpha_{\rm{SL}}=0.1,\varepsilon_{\rm{SL}}=0.1$
\State Initialize $\bm{\omega}^{(k)}_{n}$ as random configuration
\Repeat
\State Compute $\nabla_J(\bm{\omega}^{(k)}_{n})$ in~\eqref{eq:grad_J}, using~\eqref{eq:part_der}
\State Update $\bm{\omega}^{(k)}_{n}$ using~\eqref{eq:grad_ascent} with converging step size, $\mu$
\Until Convergence
\end{algorithmic}
\end{algorithm}

\section{RIS Design \& Deployment}\label{sec:ris_design}
Each RIS needs to constructively reflect the signal for any UE in $\mathcal{R}^{(k)}_{n}$, while signals from other sub-regions are suppressed. For some configuration $\bm{\omega}$, the metric
\begin{align}
\bm{H}_{k,q}\text{diag}\{\bm{\omega}\}\bm{\alpha}_{k,q}(\bm{p}) = \bm{\Lambda}_{k,q}(\bm{p})\bm{\omega}, 
\end{align}
is proportional to the received energy, via the $k^{\rm{th}}$ RIS, from a potential UE at point $\bm{p}$ and $\bm{\Lambda}_{k,q}(\bm{p}) = \bm{H}_{k,q}\text{diag}\{\bm{\alpha}_{k,q}(\bm{p})\}$. The design problem is formulated as
\begin{align}\label{eq:des:prob}
&\bm{\omega}^{\ast(k)}_{n} = \argmax_{\bm{\omega}} \sum_{\bm{p}}\sum_{q=1}^{Q} \bigl\| \bm{\Lambda}_{k,q}(\bm{p})\bm{\omega} \bigr\|^{2} , \\
&\text{such that:} |\omega_{i}| = 1 \text{ and } \sum_{\bar{\bm{p}}}\sum_{q=1}^{Q} \bigl\| \bm{\Lambda}_{k,q}(\bar{\bm{p}})\bm{\omega} \bigr\|^{2} < \varepsilon \notag ,
\end{align}
where the sum over $\bm{p}$ (sum over $\bar{\bm{p}}$) refers to some set of sampled points inside (outside) of the corresponding sub-region. Without loss of generality, let us sample the set of points corresponding to the center of mass of each sub-region. Note that~\eqref{eq:des:prob} requires the true NF channels in~\eqref{seq:channel_br} (BS-to-RIS) and a set of sampled points in $\mathcal{R}$, inspected via the NF steering vector in~\eqref{eq:alpha_NF}. Thus, the RISs are configured around the spatially dependant NF channels, creating a point-focused beampattern. Furthermore, the first constraint in~\eqref{eq:des:prob} relates to the passive nature of the RIS, while the second restricts the integrated side-lobe level below a desired threshold, $\varepsilon$. The design problem in~\eqref{eq:des:prob} is non-convex and there may be no feasible point satisfying the second condition. As a result, a relaxed design problem needs to be considered. Let us define the continuous and differentiable objective function
\begin{align}\label{eq:obj_fun}
J_{n}^{(k)}(\bm{\omega}) =& \sum_{\bm{p}}\sum_{q=1}^{Q} \bigl\| \bm{\Lambda}_{k,q}(\bm{p})\bm{\omega} \bigr\|^{2} - f_{\rm{ps}}(\bm{\omega})\\
& - \sum_{\bar{\bm{p}}}f_{\rm{SL}}\left(\sum_{q=1}^{Q} \bigl\| \bm{\Lambda}_{k,q}(\bar{\bm{p}})\bm{\omega} \bigr\|^{2}\right) , \notag
\end{align}
where the passive RIS penalty function, $f_{\rm{ps}}(\bm{\omega})$, is defined as 
\begin{align}\label{eq:pen_fun_ps}
f_{\rm{ps}}(\bm{\omega}) = \beta_{\rm{ps}}\sum_{i=1}^{N_{\rm{RIS}}}\left(|\omega_{i}| - 1 \right)^{2},
\end{align}
while the side lobe penalty function, $f_{\rm{SL}}(\chi)$, is defined as
\begin{align}\label{eq:pen_fun_sl}
f_{\rm{SL}}(\chi) = \frac{\beta_{\rm{SL}}}{1 + e^{-\alpha_{\rm{SL}}(\chi-\varepsilon_{\rm{SL}})}},
\end{align}
with $\beta_{\rm{SL}}$, $\alpha_{\rm{SL}}$ and $\varepsilon_{\rm{SL}}$ being design parameters. Essentially, \eqref{eq:pen_fun_ps} penalises any non-unit element configuration, while~\eqref{eq:pen_fun_sl} penalises any side-lobes exceeding a threshold (sigmoid activation). The gradient of the objective function in~\eqref{eq:obj_fun} is  
\begin{align}\label{eq:grad_J}
\nabla J_{n}^{(k)}(\bm{\omega}) = \left[\frac{\partial J_{n}^{(k)}(\bm{\omega})}{\partial\omega_{1}} \quad \dots \quad \frac{\partial J_{n}^{(k)}(\bm{\omega})}{\partial\omega_{N_{\rm{RIS}}}}  \right]^{T},
\end{align}
where the $i^{\rm{th}}$ partial derivative of~\eqref{eq:obj_fun} is given by 
\begin{multline}\label{eq:part_der}
\frac{\partial J_{n}^{(k)}(\bm{\omega})}{\partial\omega_{i}} = 2\sum_{\bm{p}}\sum_{q=1}^{Q}A_{k,q,i}(\bm{\omega},\bm{p}) -2\frac{\beta_{\rm{ps}}\left(|\omega_{i}| - 1 \right)\omega_{i}}{|\omega_{i}|}\\
- 2\sum_{\bar{\bm{p}}}\sum_{q=1}^{Q} f'_{\rm{SL}}\left(\sum_{q=1}^{Q} \bigl\| \bm{\Lambda}_{k,q}(\bar{\bm{p}})\bm{\omega} \bigr\|^{2}\right)\times A_{k,q,i}(\bm{\omega},\bar{\bm{p}}) ,
\end{multline}
where we define $A_{k,q,i}(\bm{\omega},\bm{p}) = \sum_{b}^{N_{\rm{BS}}} [\bm{\Lambda}_{k,q}(\bm{p})]_{i,b}[\bm{\Lambda}_{k,q}(\bm{p})\bm{\omega}]_{b}$ and $f'_{\rm{SL}}(\chi)$ is the derivative of~\eqref{eq:pen_fun_sl}, given by
\begin{align}
f'_{\rm{SL}}(\chi) = \frac{d f_{\rm{SL}}(\chi)}{d \chi} =\frac{\alpha_{\rm{SL}} \beta_{\rm{SL}} e^{-\alpha_{\rm{SL}}(\chi-\varepsilon_{\rm{SL}})}}{1+e^{-\alpha_{\rm{SL}}(\chi-\varepsilon_{\rm{SL}})}}.
\end{align}
Maximizing~\eqref{eq:obj_fun} over $\bm{\omega}$ can be carried out through the gradient ascent method, i.e., for the $\ell^{\rm{th}}$ ascent iteration we have
\begin{equation}\label{eq:grad_ascent}
\bm{\omega}_{[\ell]}=\bm{\omega}_{[\ell-1]} + \mu_{[\ell]} \nabla_J(\bm{\omega}_{[\ell-1]}),
\end{equation}
where $\mu$ is the converging gradient ascent step. The process is presented in Alg.~\ref{alg:omega_design}, for each sub-region $\mathcal{R}^{(k)}_{n}$.

RISs can be introduced in various practical scenarios, augmenting the system’s sensing abilities and offering new degrees of freedom. In particular, each RIS does not need to inspect the whole region, offering flexibility and favourable points of inspection. For example, consider the UE placement across the x-y plane. The wall-mounted BS requires high angular resolution (azimuth) and range resolution to distinguish different UEs. On the other hand, the ceiling-mounted RIS requires high resolution in the angular domain only (azimuth and elevation). Each RIS needs to have a strong LOS link with the BS and the small area it inspects, directly underneath it. The range resolution is proportional to the available bandwidth (BW), while the angular resolution is proportional to the number of antennas/RIS elements. For indoor scenarios, the BW requirements for sufficient range resolution could be unrealistically high, as indicated in Table~\ref{tab:req}. Moreover, the number of RIS elements dictates the shape of the RIS reflect-beampattern and the achievable side-lobe suppression. Thus, the RIS design that achieves proper RIS inspection requires that the considered RISs are sufficiently large for the desired performance level. Essentially, the range, azimuth and elevation resolution are derived as the inspected region is translates into polar co-ordinates for the respective RIS or BS inspection.

\begin{table}
\caption{Considered geometry \& resolution requirements}
\label{tab:req}
\begin{center}
\renewcommand{\arraystretch}{1.5} 
\begin{subtable}[t]{\textwidth}
\begin{tabular}{| m{8.5em} | m{18.95em}|} 
    \hline
    \multicolumn{2}{|c|}{\cellcolor{lightgray}{RIS placement \& Considered region: $K_{\rm{RIS}} = 3$}} \\
    \hline	
    $\;\bm{p}_{\rm{BS}} = [4.5, 0, 2]       $ & $\quad\;\mathcal{R}:\rm{x}\in[0,9], \rm{y}\in[3:6], \rm{z}\in[1.4:1.8] $\\
    $\bm{p}_{\rm{RIS}}^{(1)} = [1.5, 4.5, 3]$ & $\mathcal{R}^{(1)}:\rm{x}\in[0,3], \rm{y}\in[3:6], \rm{z}\in[1.4:1.8]  $\\
    $\bm{p}_{\rm{RIS}}^{(2)} = [4.5, 4.5, 3]$ & $\mathcal{R}^{(2)}:\rm{x}\in[3,6], \rm{y}\in[3:6], \rm{z}\in[1.4:1.8]  $\\
    $\bm{p}_{\rm{RIS}}^{(3)} = [7.5, 4.5, 3]$ & $\mathcal{R}^{(3)}:\rm{x}\in[6,9], \rm{y}\in[3:6], \rm{z}\in[1.4:1.8]  $\\
    \hline
\end{tabular}
\end{subtable}
\begin{subtable}[h]{\textwidth}
\begin{tabular}{| m{29em}|}
    \hline
    \multicolumn{1}{|c|}{\cellcolor{lightgray}{Requirements for RIS aided inspection: $\mathcal{R}^{(k)}_{n} = 0.3\times0.3\times0.4\text{ m}^{3}$}} \\
    \hline	
    Azimuth resolution: $\Delta\theta^{\circ} = 6.28^{\circ} \rightarrow N_{\rm{RIS}}^{(\rm{x})} \geq 15>\frac{0.8\lambda_{\rm{o}}}{\delta_{\rm{RIS}}\Delta\theta^{\circ}}\frac{180}{\pi} \;\;\;\;$\\ 
    Elevation resolution: $\Delta\theta^{\circ} = 6.28^{\circ} \rightarrow N_{\rm{RIS}}^{(\rm{y})} \geq 15>\frac{0.8\lambda_{\rm{o}}}{\delta_{\rm{RIS}}\Delta\theta^{\circ}}\frac{180}{\pi}$         \\ 
    \hline
    \multicolumn{1}{|c|}{\cellcolor{lightgray}{Requirements for direct BS inspection: $\mathcal{R}^{(k)}_{n} = 0.3\times0.3\times0.4\text{ m}^{3}$}} \\
    \hline   
    Azimuth resolution: $\Delta\theta^{\circ} = 2.72^{\circ} \rightarrow N_{\rm{BS}}^{(\rm{x})} \geq 34>\frac{0.8\lambda_{\rm{o}}}{\delta_{\rm{BS}}\Delta\theta^{\circ}} \frac{180}{\pi}$\\ 
    Range  resolution: $\Delta\rm{d} = 0.3 $[m]$\rightarrow $ BW $\geq 1$ GHz $= \frac{c_{\rm{o}}}{\Delta\rm{d}}$\\ 
    \hline
    \multicolumn{1}{|c|}{\cellcolor{lightgray}{Requirements for direct BS inspection: $\mathcal{R}^{(k)}_{n} = 1\times1\times0.4\text{ m}^{3}$}} \\
    \hline 
    Azimuth resolution: $\Delta\theta^{\circ} = 6.61^{\circ} \rightarrow N_{\rm{BS}}^{(\rm{x})} \geq 14>\frac{0.8\lambda_{\rm{o}}}{\delta_{\rm{BS}}\Delta\theta^{\circ}} \frac{180}{\pi}$\\ 
    Range  resolution: $\Delta\rm{d} = 1 $[m]$\rightarrow $ BW $\geq 300$ MHz $= \frac{c_{\rm{o}}}{\Delta\rm{d}}$\\ 
    \hline
\end{tabular}
\end{subtable}
\end{center}
\end{table}

\section{UE detection}
In this section, we exploit the non-static signal component over different frames for UE detection. The RIS-assisted link is designed to produce a combination of frames that depend on the static channel exclusively, i.e., when certain known RIS configurations are combined,  only the LOS and the static multipath link is observed by the BS. The transmitted pilots and the corresponding RIS configurations can be designed such that there is an $\tilde{n}$ where $\dot{\bm{y}}_{q,n} = 1/2\bigl(\bm{y}_{q,n}+\bm{y}_{q,\tilde{n}}\bigr)$ and
\begin{align}\label{LOS_NLOS_sep}
\dot{\bm{y}}_{q,n} = \sqrt{P_{\rm{sym}}} \sum_{m=1}^{M} \dot{\bm{h}}_{q}^{(m)}\otimes \bm{x}_{q,n}^{(m)}  + \dot{\bm{z}}_{q,n} \; . 
\end{align}
The procedure to make such a design is not trivial. The coherence time of the channel, the cost of having multiple transmissions and the overall desired application are some of the factors that may shape the RIS design process. Here, we consider that the UEs transmit $\tilde{N}$ frames separately, for this procedure. However, this is not a necessity, and is a strategy to simplify notations. The frames designed to aid with with the signal extraction may need to be part of a more complex transmission procedure. A thorough investigation of the pilot design and transmission strategy that achieves~\eqref{LOS_NLOS_sep} is left for future consideration. Considering the case with $\tilde{N}=N$, a simple design is to set $\tilde{n}=n-\tilde{N}$ and
\begin{align}
\bm{x}_{q,n}^{(b)} = \bm{x}_{q,\tilde{n}}^{(b)}  , \quad \bm{\omega}_{\tilde{n}}^{(k)} = -\bm{\omega}_{n}^{(k)} , \quad \forall n&=1,\dots,N  .
\label{eq:split_cond}
\end{align}
Thus, we can utilize~\eqref{LOS_NLOS_sep} to isolate the non-static component of the signal, as $\bar{\bm{y}}_{q,n} = \bm{y}_{q,n} - \dot{\bm{y}}_{q,n}$, where
\begin{align}
\label{y_bar}
\bar{\bm{y}}_{q,n} = \sqrt{P_{\rm{sym}}} \sum_{m=1}^{M} \bar{\bm{h}}_{q,n}^{(m)}\otimes \bm{x}_{q,n}^{(m)} + \bar{\bm{z}}_{q,n} \; . 
\end{align}
The energy received via the RIS's reflection changes over time, as the RIS uses different configurations, employing an iterative scanning procedure. The combination of all RISs' contributions in~\eqref{y_bar} can be plugged into

\begin{align}
f_{n}^{(b,k)} = \biggl | \sum_{q=1}^{Q} \biggl(\bm{t}_{q,n}^{(k)}(\bm{\omega}_{n}^{(k)}) \otimes \bm{x}_{q,n}^{(b)}\biggr)\herm \bar{\bm{y}}_{q,n} \biggr | \; ,
\label{eq:matched_filter}
\end{align}
for $b=1,\dots,B$, and $n=1,\dots,N$. The output of~\eqref{eq:matched_filter} is essentially a filter matched to the cascaded channel for the point-wise inspection via the corresponding RIS, for each individual RB, producing a total of $KB$ sequences of length $N$. Each sequence, $\bm{f}^{(b,k)} = [f_{1}^{(b,k)},\dots,f_{N}^{(b,k)}]\transp \in \mathbb{R}^{N\times 1}$ is connected with the region $\mathcal{R}^{(k)}$ inspected by the RIS and it tracks how much energy is reflected by the $k^{\rm{th}}$ RIS over the $N$ OFDM frames. If $\bm{\omega}_{n}^{(k)}$ is designed according to Alg.~\ref{alg:omega_design}, to inspect $\mathcal{R}_{n}^{(k)}$, then the energy detector
\begin{align}
f_{n}^{(b,k)} \underset{\mathcal{H}_{0}}{\overset{\mathcal{H}_{1}}{\gtrless}} \gamma \left\{\begin{matrix}
\mathcal{H}_{1}: \exists \text{ UE}^{(m^{\ast})} \in \mathcal{R}_{n}^{(k)} \text{ s.t.:} \bm{x}^{(m^{\ast})} \equiv \bm{x}^{(b)}
\\ 
\mathcal{H}_{0}:\nexists \text{ UE}^{(m^{\ast})} \in \mathcal{R}_{n}^{(k)} \text{ s.t.:} \bm{x}^{(m^{\ast})} \equiv \bm{x}^{(b)}
\end{matrix}\right.
\label{eq:en_det}
\end{align}
indicates the presence of a UE, $m^{\ast}$, within the inspected region, $\mathcal{R}_{n}^{(k)}$, using the $b^{\rm{th}}$ available pilot. In~\eqref{eq:en_det}, if the received energy exceeds the set threshold $\gamma$, hypothesis $\mathcal{H}_{1}$ where there is a UE in $\mathcal{R}_{n}^{(k)}$ is assumed to be true. Otherwise, the null hypothesis, $\mathcal{H}_{0}$, of no present UEs is true. Essentially, the energy detector in~\eqref{eq:en_det} augments the traditional concept of signal detection to incorporate the spatial information of the corresponding signal. The corresponding spatial information is extracted from the scanning procedure, undertaken by the RISs.

\textbf{Complexity Analysis}: The complexity of our proposed detection procedure scales with the desired inspection resolution. In particular, there are $NK$ sub-regions, and each one requires $B$ iterations of~\eqref{eq:matched_filter} for each possible pilot, i.e., $N$ OFDM frames and $NKB \times QLN_{\rm{BS}}$ complex multiplications are required. Note that the computational complexity scales linearly with each parameter and no exhaustive search is required, in contrast to many traditional localization algorithms~\cite{10621044}.

\section{Channel Access \& Detection}
In this section, we examine, in closed-form expressions, the effect of channel collisions in UE detection. As the undetected UEs access the channel randomly, i.e., they use a random RB, the number of UEs that access an RB essentially follows a binomial distribution. The probability that $m$ UEs share access of the same RB is given by
\begin{align}
\mathbb{P}\{m:M,B_{\rm{R}} \} = \binom{M}{m} \biggl(\frac{1}{B_{\rm{R}}} \biggr)^{m} \biggl(\frac{B_{\rm{R}} -1}{B_{\rm{R}}} \biggr)^{(M - m)},
\label{eq:pdf_m_M_B}
\end{align}
for $m = \{1,\dots,M \}$, where $M$ and $B_{\rm{R}}$ are the total number of active UEs and the number of RBs available.

The probability that a UE can be detected depends on the RB congestion and the transmit power. A UE can be detected if $M$ total UEs transmit with $\mathcal{P}_{\rm{sym}}$ over $B_{R}$ RBs with a probability of
\begin{align}
\mathbb{P}&
\{\text{detection}:M,B_{\rm{R}},\mathcal{P}_{\rm{sym}} \} = 
\label{eq:p_det}
\\ &\sum_{m=1}^{M} \mathbb{P}\{m:M,B_{\rm{R}}\}\times
\mathbb{P}\{\rm{detection}:m,\mathcal{P}_{\rm{sym}} \} \; , \notag
\end{align}
where $\mathbb{P}\{\text{detection}:m,\mathcal{P}_{\rm{sym}} \}$ denotes the probability of detection when $m$ UEs share a single RB and it depends, among other factors, on the considered geometry. Therefore, the probability that $m$ UEs are detected is
\begin{align}
\mathbb{P}&\{\rm{detections} = m:M,B_{\rm{R}},\mathcal{P}_{\rm{sym}} \} = 
\\ & \binom{M}{m} \biggl(1-\mathbb{P}\{\rm{detection}:M,B_{\rm{R}},\mathcal{P}_{\rm{sym}} \}\biggr)^{(M -m)} \notag 
\\ & \times  \biggl(\mathbb{P}\{\rm{detection}:M,B_{\rm{R}},\mathcal{P}_{\rm{sym}} \} \biggr)^{m} \; . \notag
\end{align}
If we assume no newly active UEs and the procedure is repeated over multiple detection phases, say $J$, the number of UEs that access the channel randomly may decrease, using an adaptive allocation strategy, i.e., the BS may assign specific RBs to detected UEs. Assuming that the transmit power and the number of available pilots remain constant, we derive the simpler notation $\mathbb{P}\{m:M_{j}\}$ to mark the probability that $m$ out of $M_{j}$ undetected UEs are detected during the $j^{\rm{th}}$ detection phase. After $J$ phases, the discrete probability distribution of the total number of detected UEs is given by
\begin{align}
\bigl[\mathbf{p}_{M}^{(J)}\bigr]_{m} &= \sum_{i=0}^{M}\sum_{k=0}^{M} \bigl[\tilde{\mathbf{p}}_{M}^{(J)}\bigr]_{i} \times \bigl[\mathbf{p}_{M}^{(J-1)}\bigr]_{k} \; , i+k=m,
\end{align}
for $m = 0,\dots ,M$, where the vector $\tilde{\mathbf{p}}_{M}^{(J)}$ gives the probability of new UEs being detected and it is given by
\begin{align}
\tilde{\mathbf{p}}_{M}^{(J)} = \mathbf{p}_{M}^{(J-1)} \times \mathbf{P}_{M} \in \mathbb{R}^{1\times(M+1)}\; ,
\end{align}
with $\mathbf{p}_{M,J-1}$ denoting the probability distribution of the previous phase and the matrix
\begin{align}
\mathbf{P}_{M} &= \begin{bmatrix}
\mathbb{P}\{0:M\} & \mathbb{P}\{1:M\} & \dots & \mathbb{P}\{M:M\} \\ 
\mathbb{P}\{0:M-1\} & \mathbb{P}\{1:M-1\} & \dots & 0 \\ 
\dots & \dots & \dots & \dots \\ 
\mathbb{P}\{0:1\} & \mathbb{P}\{1:1\} & \dots & 0 \\ 
1 & 0 & \dots & 0 \\ 
\end{bmatrix} \notag 
\\ & \quad\quad\quad\quad\quad\quad\quad\quad\quad\quad\quad \in \mathbb{R}^{(M+1)\times(M+1)} \; .
\end{align}
accounts for different probability distributions for new UEs being detected, based on the total number of undetected UEs. The initial distribution ($J=1$) is simply
\begin{align}
\mathbf{p}_{M}^{(1)} = \bigl[\mathbb{P}\{0:M\}, \dots , \mathbb{P}\{M:M\} \bigr] \in \mathbb{R}^{1\times(M+1)} \; .
\end{align}

\section{Simulation Results}\label{sec:sim}
Consider Fig.~\ref{fig:sm} and the simulation parameters given in Table~\ref{tab:sim_par}. We consider the 6 GHz band, as one of the 5G bands. Other frequency bands are also applicable. For the considered geometry, the Fraunhoffer distance is at 29.78 m (28.78 m) for the BS (RIS). Any propagation path is much smaller for the considered geometry, highlighting the validity of the NF assumption. However, our core idea can be extended to an FF scenario, if the RIS configurations are designed accordingly. The element gain is modelled as $G(\bm{\theta}) = \cos^{2}(\theta_{\rm{az}})\cos^{2}(\theta_{\rm{el}})$. 

\begin{table}
\caption{Simulation Parameters}
\label{tab:sim_par}
\begin{center}
\renewcommand{\arraystretch}{1.5} 
\begin{tabular}{|l  l | l  l | l l |}
    \hline
    \multicolumn{6}{|c|}{\cellcolor{lightgray}{Simulation Parameters}} \\
    \hline	
    $f_{\rm{o}}$                & 6 GHz  & $W_{\rm{o}}$                & 30 KHz  & $Q$                   & 1                     \\
    $N_{\rm{BS}}^{(\rm{x})}$    & 34     & $N_{\rm{BS}}^{(\rm{y})}$    & 6       & $\delta_{\rm{BS}}$    & $\lambda_{\rm{o}} /2$ \\
    $N_{\rm{RIS}}^{(k,\rm{x})}$ & 24     & $N_{\rm{RIS}}^{(k,\rm{y})}$ & 24      & $\delta_{\rm{RIS}}$   & $\lambda_{\rm{o}} /2$ \\
    BW                          & 15 KHz & $L$                         & 1       & $B$                   & 5                     \\
    $N$                         & 100    & $\rm{Noise}_{\rm{F}}$       & 10 dB   & $\rm{Noise}_{\rm{D}}$ & -174 dBm/Hz           \\
    \hline
\end{tabular}
\end{center}
\end{table}

\begin{figure}[h]
    \centering
    \includegraphics[width=\columnwidth]{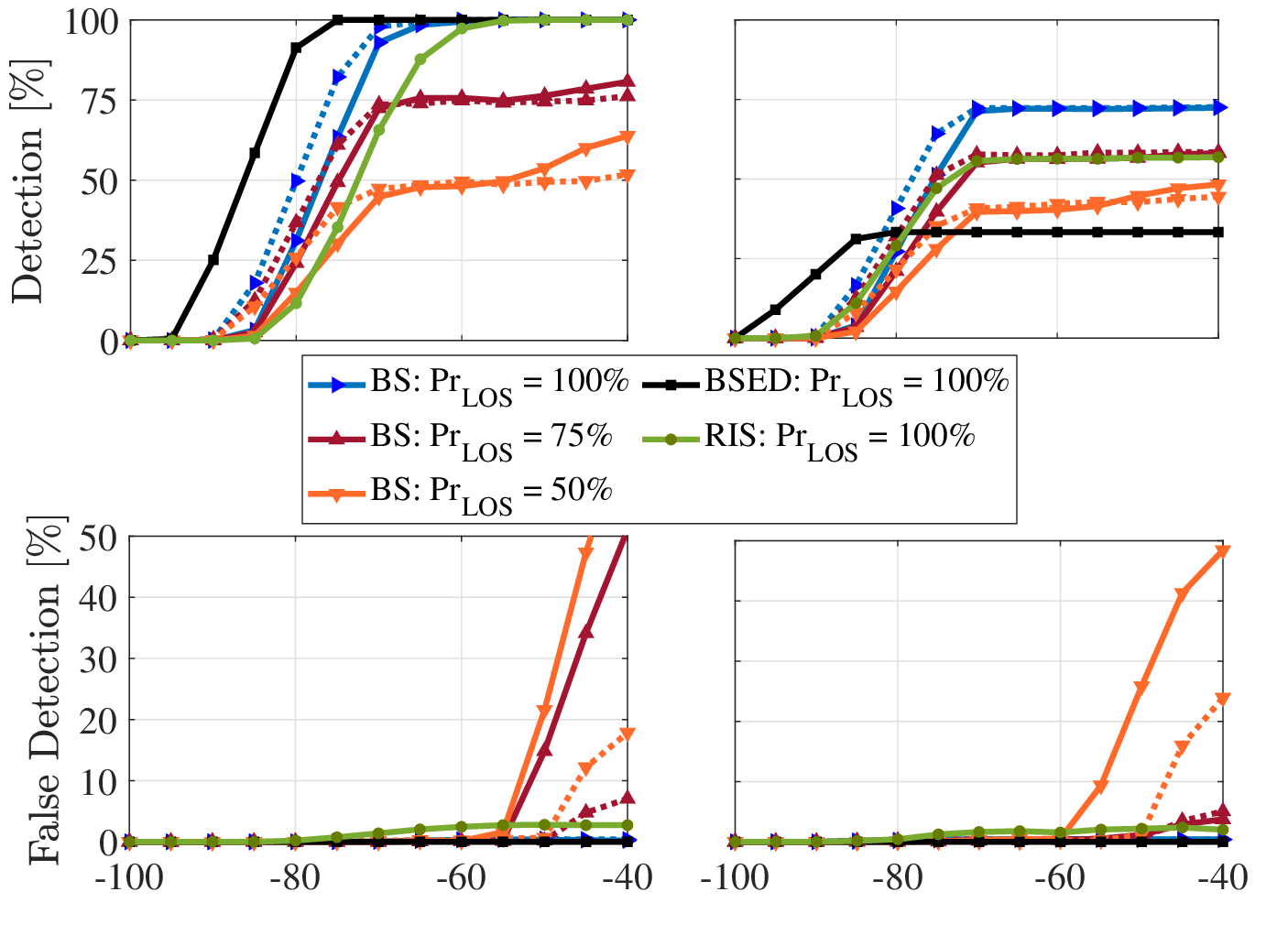}
    \caption{Performance with and without RIS, for a single active UE per RB (left, $M=1$) or multiple (right, $M=3$). The solid / dotted lines indicate strong $(K_{\rm{Rice}} = 1)$ / weak multipath $(K_{\rm{Rice}} = 10)$, respectively.}
    \label{fig:no_ris}
\end{figure}

Figure~\ref{fig:no_ris} compares the performance of a system operating with or without RISs, either for a single (left) or multiple (right) active UEs. As illustrated in Table~\ref{tab:req}, direct inspection by the BS requires unrealistically large BW to achieve the desired resolution. The total channel BW for the BS inspection is set to $1.188$ GHz ($Q = 100, W_{\rm{sub}} = 12$ MHz), as opposed to the case with RISs in Table~\ref{tab:sim_par} (BW = $30$ KHz). As the LOS probability and the multipath level vary, the simulations show that a strong LOS is imperial for the performance of such a system, as the probability of detection experiences saturation. For perfect LOS conditions, the BS inspection seems to outperform the RIS-based system due to the stronger direct link. However, this comes with the cost of significantly larger BW. 

When multiple UEs are active within an RB, the inter-user interference significantly affects our ability to properly detect active UEs and estimate the position of the signal source, regardless of the inspection method and the considered resources. The performance degradation is more prominent when no RISs are considered, and the LOS link is obstructed. 

If the BW is not increased, the UE detection may be reduced to a standard energy detection (ED) problem, as considered in~\cite{li2022joint} (black line), i.e., there is no RIS, and no position information is extracted. Essentially, the RISs transform UE detection to spatially aware detection, without degrading the detection performance. Note the saturated probability of detection in standard ED, when multiple UEs are active with the same RB. Without the RISs, the standard ED problem completely fails to combat the multiple access problem; the BS can not distinguish when multiple UEs access an RB.

In addition, when the LOS is obstructed and no RIS is utilized (orange and yellow lines), the BS still detects the activated resource blocks when the transmitted energy is sufficiently large, through the random multipath links. However, no spatial information is extracted from the random multipath and the probability of false detection significantly increases. Lastly, the probability of false detection slightly increases for high transmit power in the RIS-aided localization, since the side lobes of the RIS response produce false alarms. However, a stricter side-lobe suppression could be incorporated in the design of the RIS configuration in Algorithm~\ref{alg:omega_design}.

\begin{figure}[h]
    \centering
    \includegraphics[width=\columnwidth]{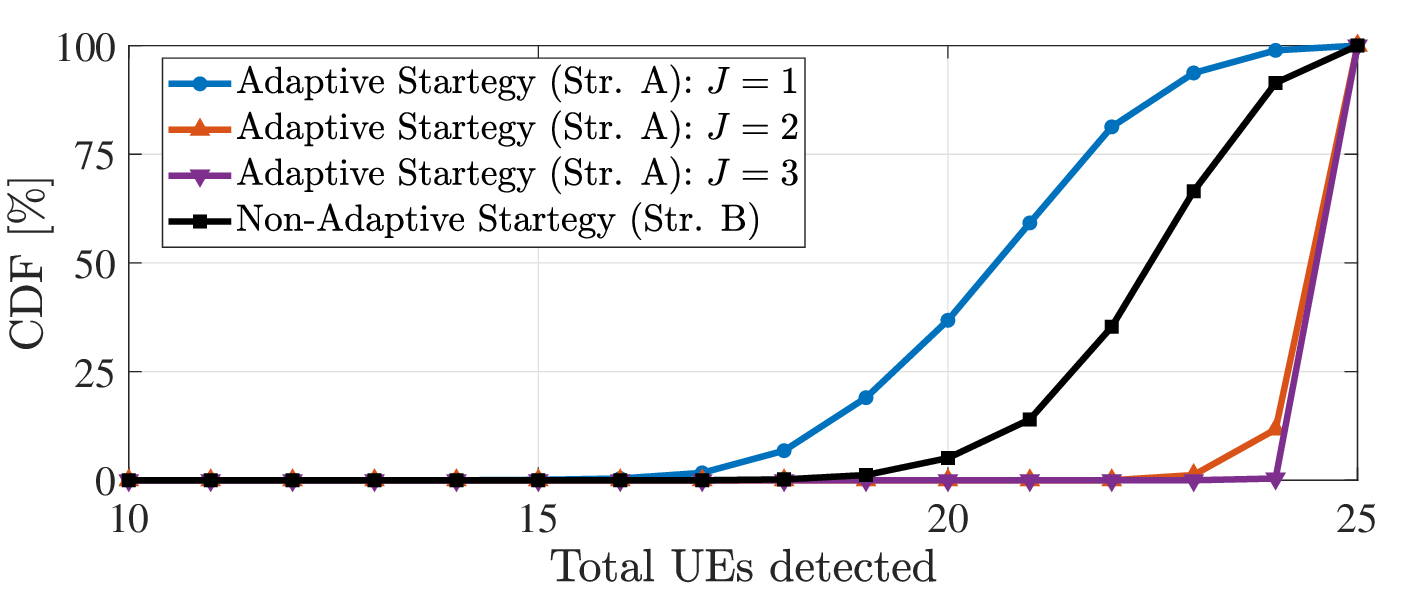}
    \caption{CDF for multiple detection phases and different resource management Strategies: Adaptive and Non-Adaptive.}
    \label{fig:cdf_J}
\end{figure}

Figure~\ref{fig:cdf_J} shows the cumulative distribution function (CDF) for the number of detected UEs, for different numbers of detection phases and for different channel access strategies ($P_{\rm{sym}} = - 50$ dBW). There are 25 active UEs and 50 available RBs. Strategy A refers to our proposed solution; a set of RBs is reserved for detected UEs ($B_{\rm{R}}=25, B_{\rm{A}}=25$). Strategy B refers to random channel access for each detection phase, without separating the detected and the undetected UEs ($B_{\rm{R}}=50, B_{\rm{A}}=0$). Strategy B allocates all RBs for the random access, making it more effective when the detection procedure is performed only once. However, for multiple iterations of the detection procedure Strategy A outperforms Strategy B, since the detected UEs no longer interfere with the undetected ones, while Strategy B does not adapt with multiple iterations. Even a single iteration is sufficient to efficiently decongest the channel access, as multiple UEs are detected and do not interfere in upcoming detections procedures. With our proposed adaptive strategy practically all the UEs are detected for more than one iteration ($J>1$). In general, multiple iterations over different RBs provide a more favourable detection environment, as the channel access is structured, and fewer collisions are expected.

\begin{figure}[h]
    \begin{subfigure}{\columnwidth}
    \centering
    \includegraphics[width=\columnwidth]{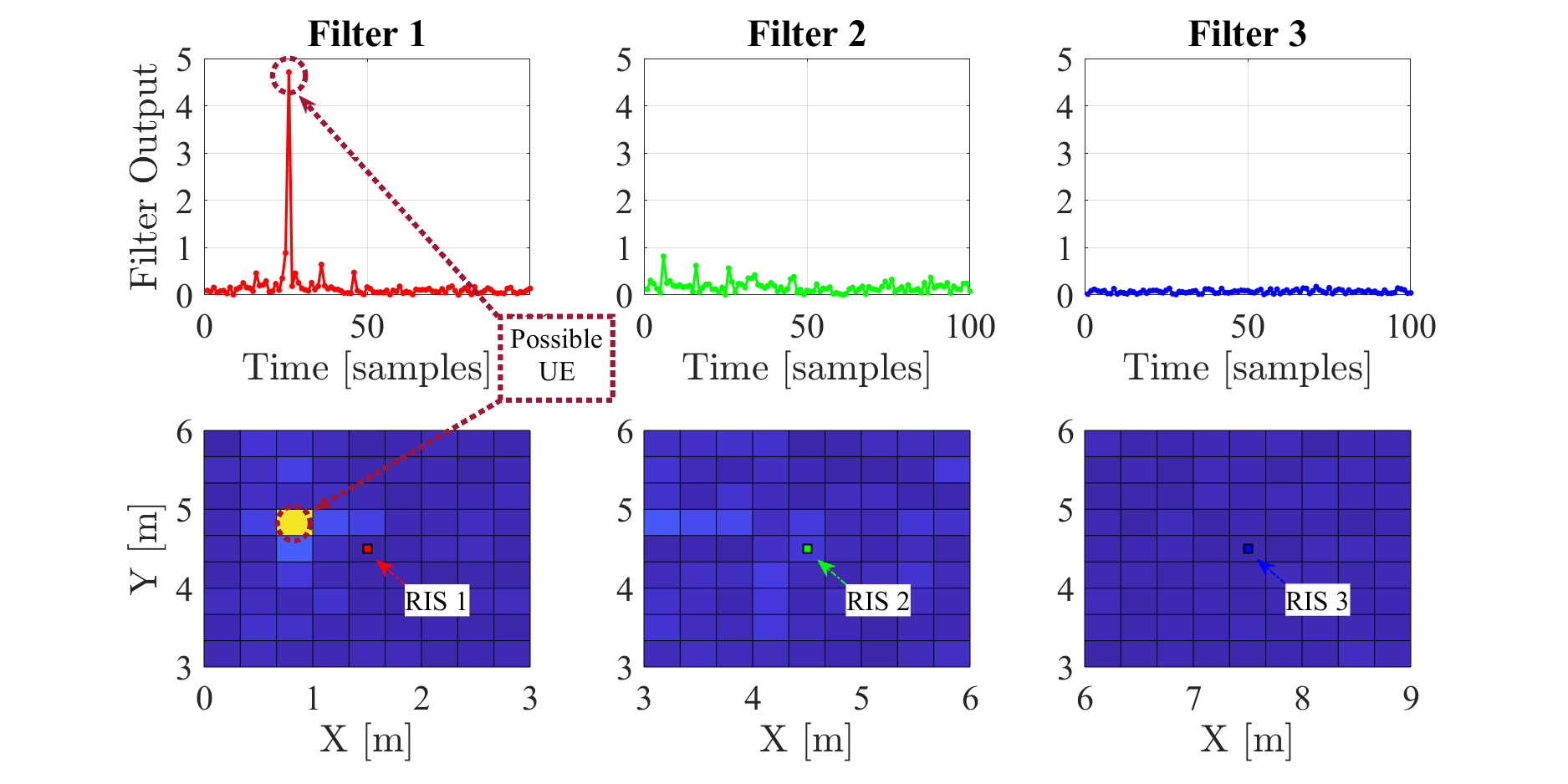}
    \caption{RB 1}
    \label{sfig:fltr_srf_1}
    \end{subfigure}
    \hfill
    \begin{subfigure}{\columnwidth}
    \centering
    \includegraphics[width=\columnwidth]{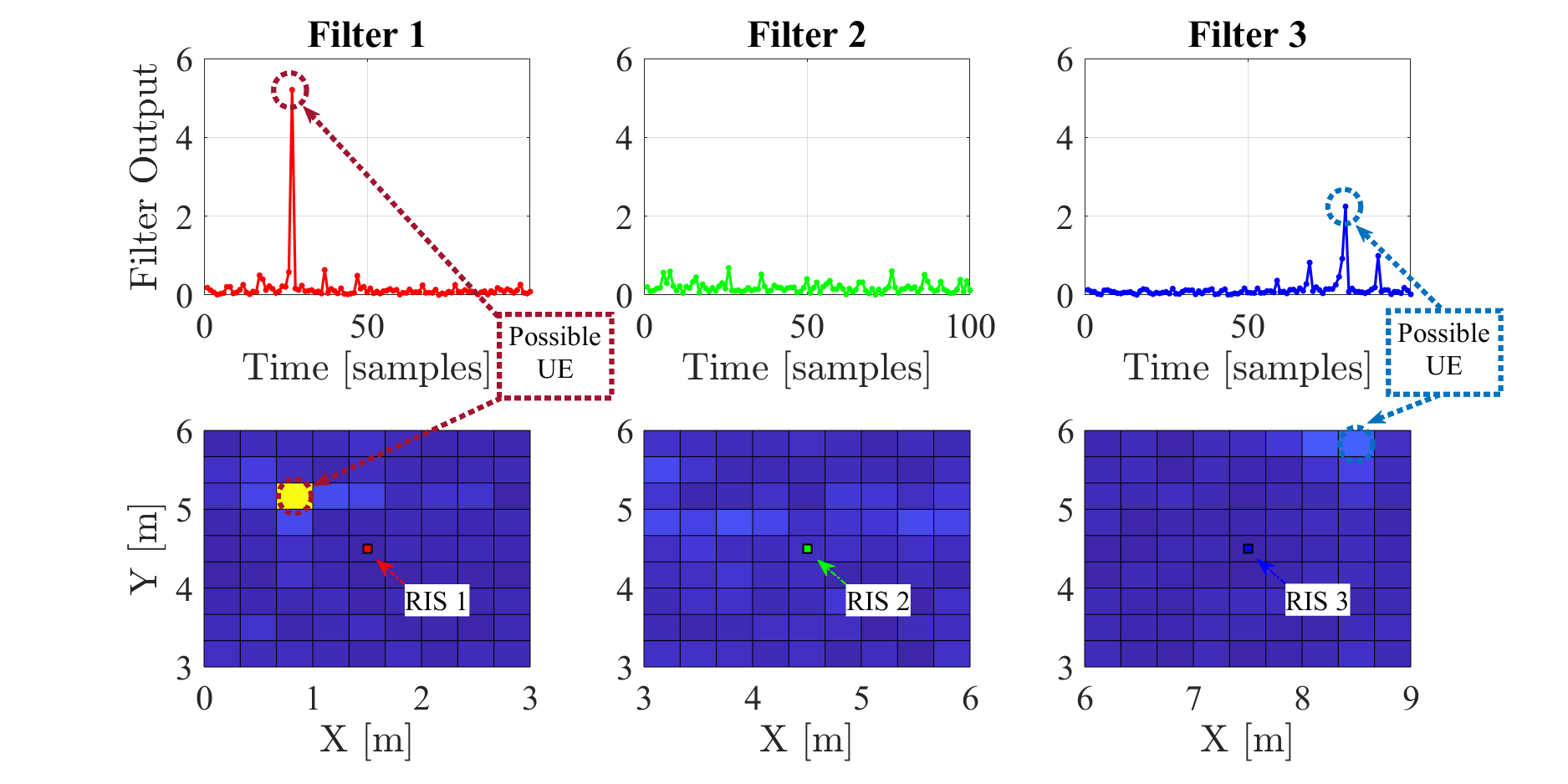}
    \caption{RB 2}
    \label{sfig:fltr_srf_2}
    \end{subfigure}
    \hfill
    \begin{subfigure}{\columnwidth}
    \centering
    \includegraphics[width=\columnwidth]{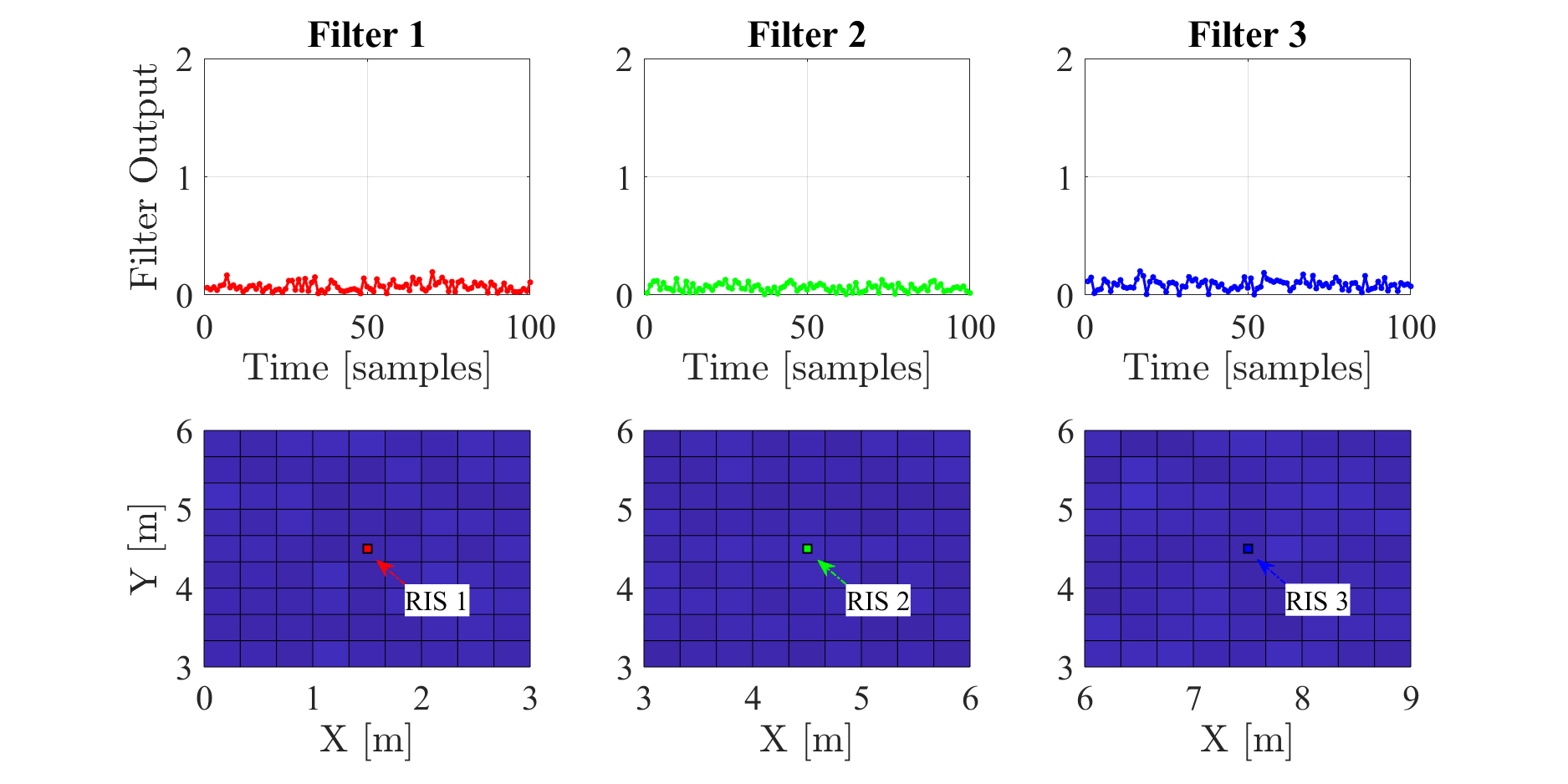}
    \caption{RB 3}
    \label{sfig:fltr_srf_3}
    \end{subfigure}
    \caption{Filter outputs in time for different RBs and the geometrical projection into $\mathcal{R}$.}
    \label{fig:fltr_srf}
\end{figure}

\begin{figure}[h]
    \begin{subfigure}{\columnwidth}
    \centering
    \includegraphics[width=\columnwidth]{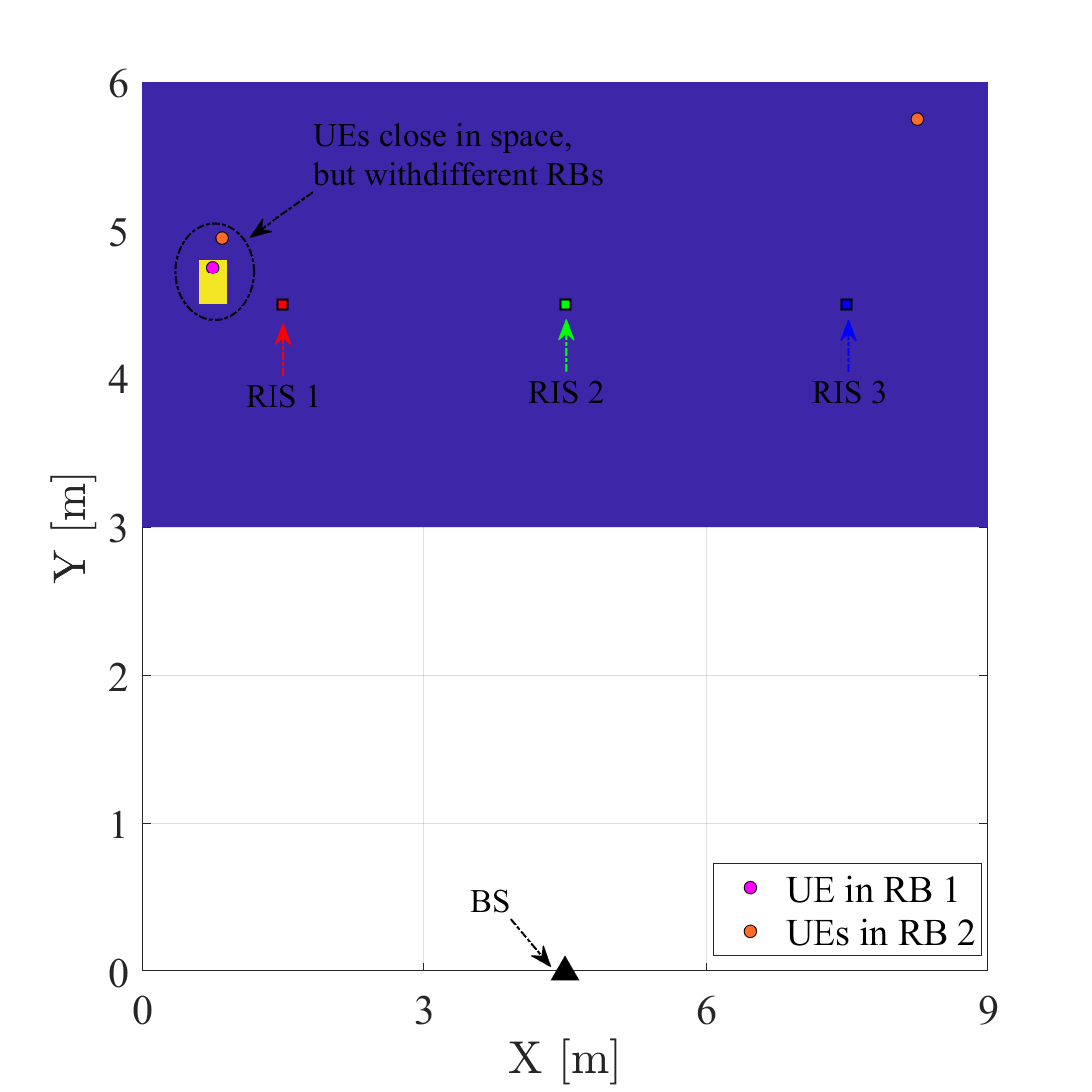}
    \caption{RB 1}
    \label{sfig:comb_srf_1}
    \end{subfigure}
    \hfill
    \begin{subfigure}{\columnwidth}
    \centering
    \includegraphics[width=\columnwidth]{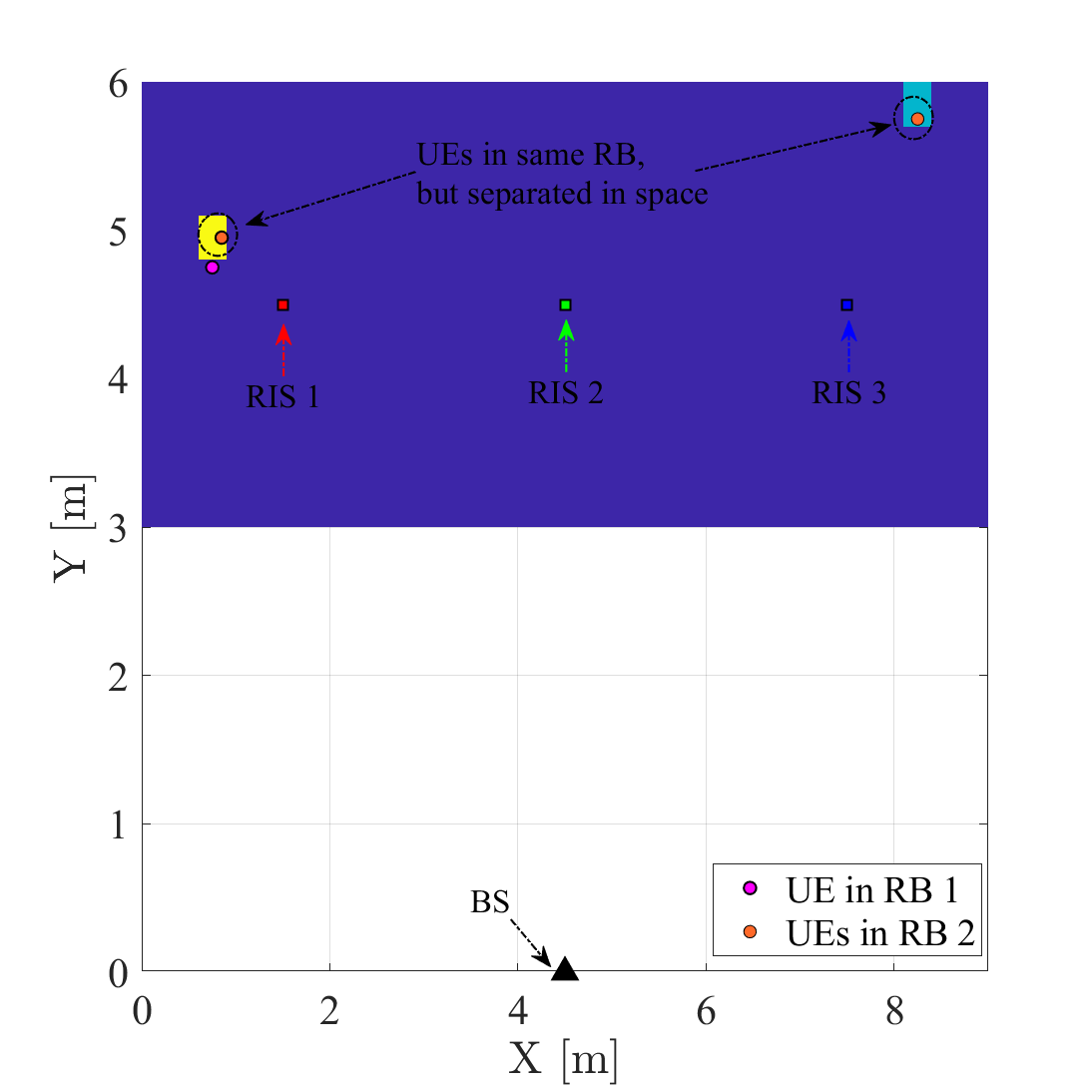}
    \caption{RB 2}
    \label{sfig:comb_srf_2}
    \end{subfigure}
    \end{figure}%
    \begin{figure}[ht]\ContinuedFloat
    \centering
    \begin{subfigure}{\columnwidth}
    \centering
    \includegraphics[width=\columnwidth]{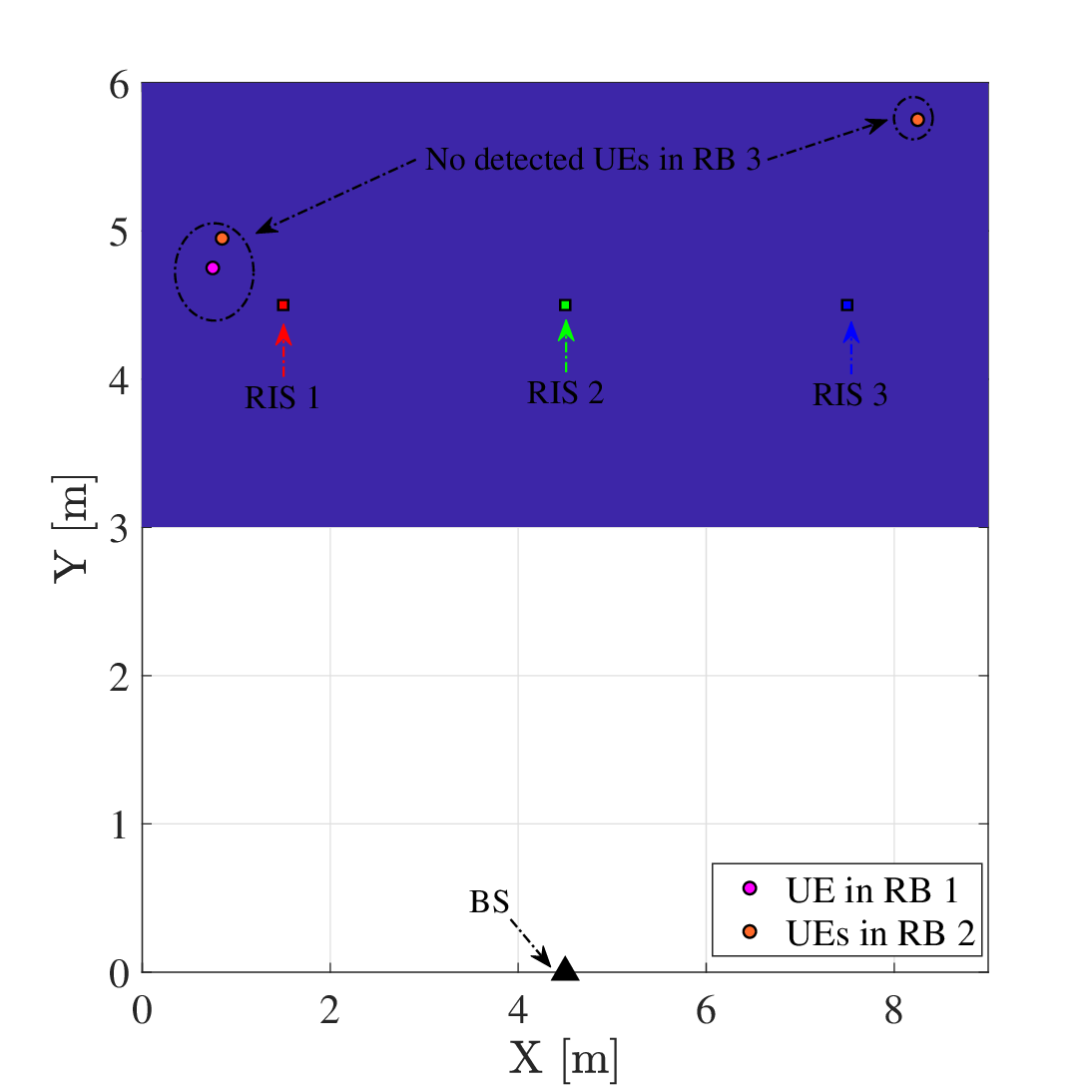}
    \caption{RB 3}
    \label{sfig:comb_srf_3}
    \end{subfigure}
    \caption{Combination of filter outputs for different RBs and the geometrical translation into $\mathcal{R}$.}
    \label{fig:comb_srf}
\end{figure}

Figures~\ref{fig:fltr_srf} and~\ref{fig:comb_srf} illustrate the sequential steps of the proposed detection scheme. We present an example with 3 UEs ($M=3$) and 3 RBs ($B_{\rm{R}}=3$), for a single iteration ($J=1$). In Fig.~\ref{fig:fltr_srf} the filter outputs are displayed across various time samples for all available RBs. Each filter is specifically designed for a corresponding RIS and focuses on a smaller inspection region. When the filter output surpasses the detection threshold, a UE is identified as having accessed the respective RB. The RIS response pinpoints the detected UE at the inspected location for that time sample. 

Figure~\ref{fig:comb_srf} showcases a composite map of the entire region by aggregating the filter outputs across all RBs. The detection protocol successfully identifies and localizes the two UEs that accessed RB 2, as their signals were received through different RISs. Additionally, the two UEs located close to each other are accurately detected and localized, since they accessed different RBs. It is important to note that no UE accessed RB 3, highlighting that the BS is not able to orchestrate a structured channel access, as the UE activity is not known.

\begin{figure}[h]
    \centering
    \includegraphics[width=\columnwidth]{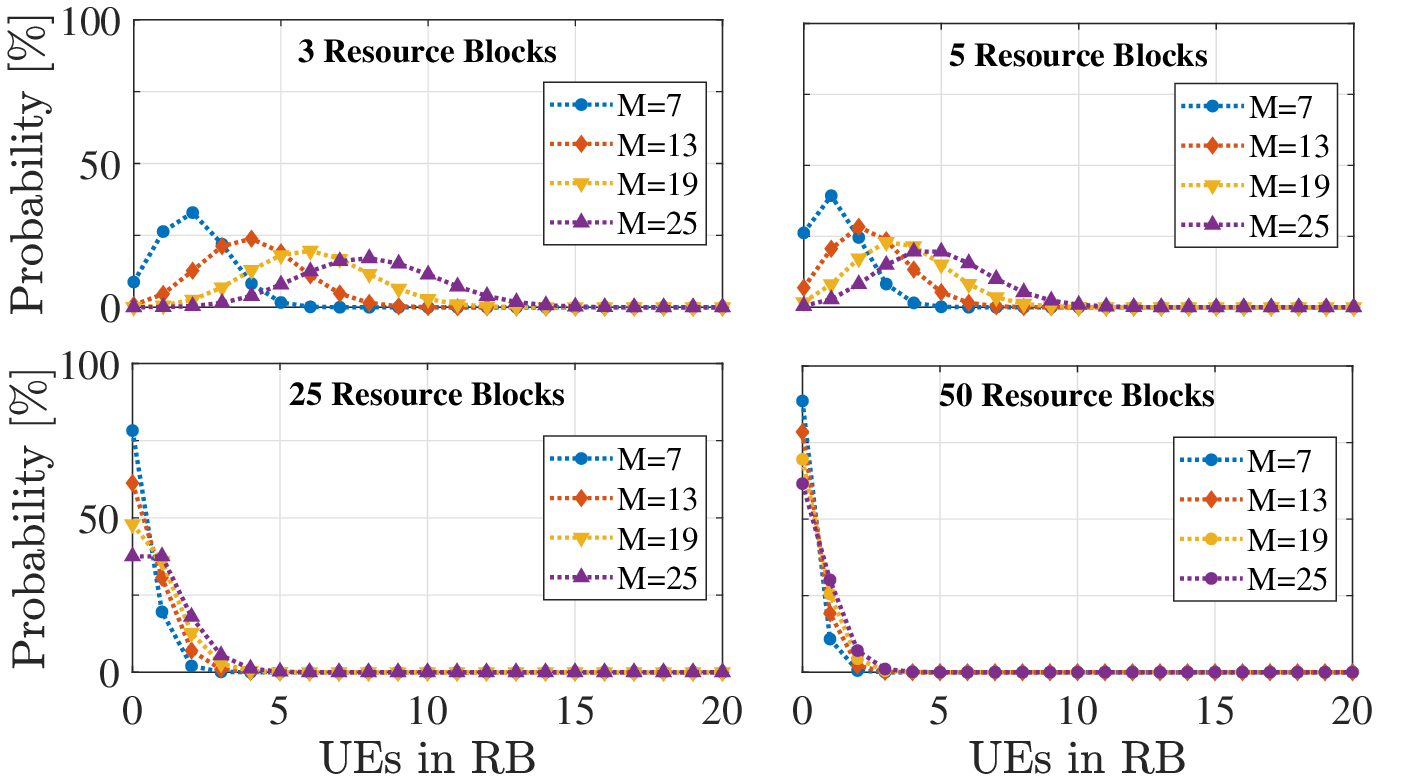}
    \caption{Probability of channel congestion for random channel access.}
    \label{fig:p_cong}
\end{figure}

Figure~\ref{fig:p_cong} illustrates the probability of collision in random channel access for different numbers of UEs and available RBs. As expected, channel congestion decreases as more RBs become available for random access. When the number of UEs is comparable to the available RBs, multiple collisions are likely. However, when the available RBs exceed the number of UEs accessing the channel, the reduction in expected collisions becomes marginal. This behavior suggests that our proposed adaptive allocation protocol is most effective in scenarios where the number of available RBs slightly exceeds the anticipated number of active UEs. Otherwise, the BS may lack the flexibility to reduce the number of random access RBs.

\section{Conclusions}
We have proposed an iterative algorithm for detection/localization of an unknown number of UEs. Our proposed adaptive detection process allows the BS to control the allocation of resources for the detected UEs and increases the probability of detection, as multiple iterations are considered. Moreover, we have shown that the restrictions on the BS placement for specific geometrical configurations lead to infeasible requirements for proper inspection, while the presence of RISs reduces the required BW significantly and improves the localization/detection performance by bypassing blockages and providing a second view on the UEs.

\bibliographystyle{ieeetr}
\bibliography{references}

\end{document}